\documentclass[paper,prd,reprint,preprintnumbers,nofootinbib,notitlepage,showpacs,showkeys]{revtex4-1}
\usepackage{latexsym,graphicx,amssymb,amsmath,mathrsfs} 

\DeclareSymbolFont{bbold}{U}{bbold}{m}{n}
\DeclareSymbolFontAlphabet{\mathbbold}{bbold}

\newcommand{\be}{\begin{equation}}      
\newcommand{\ee}{\end{equation}}      
\newcommand{\bea}{\begin{eqnarray}}      
\newcommand{\eea}{\end{eqnarray}}    
     
\newcommand{\rt}[1]{{}}

\makeatletter
\renewcommand\appendix{\par
\setcounter{section}{0}%
\setcounter{subsection}{0}%
\gdef\thesection{\appendixname\space\@Alph\c@section}}

\long\def\unmarkedfootnote#1{{\long\def\@makefntext##1{##1}\footnotetext{#1}}}
\makeatother

\begin{document} 

\title{Renormalization of the 2PI-Hartree approximation in a broken phase \\ with nonzero superflow}

\author{G. Fej\H{o}s}
\email{fejos@riken.jp}
\affiliation{Theoretical Research Division, Nishina Center, RIKEN, Wako 351-0198, Japan}

\begin{abstract} 
Nonperturbative renormalization and explicit construction of the effective potential of the Hartree approximation of the two-particle-irreducible formalism are carried out in an inhomogeneous field configuration describing a uniform superfluid. Based on the earlier article [G. Fej\H{o}s {\it et. al}, Nucl. Phys. {\bf A803}, 115 (2008)], we clarify certain aspects of renormalizability corresponding to the findings of [M. G. Alford {\it et. al}, Phys. Rev. D {\bf 89}, 085005 (2014)]. We show that renormalizability of the approximation can be ensured by regularization schemes respecting Lorentz and translation invariance. Elimination of nonconventional superflow-dependent divergences is presented in detail, together with a discussion on the finite-temperature treatment.
\end{abstract}

\pacs{11.10.Gh}
\keywords{2PI formalism, superfluidity}  
\maketitle

\section{Introduction}

The two-particle-irreducible (2PI) formalism is a popular functional method applied to quantum field theories both in and out of equilibrium. The key quantity of the formulation is the 2PI effective action \cite{cornwall74}, which contains the mean field and also the propagators as variables. Stationary conditions of the action lead to equations for the one- and two-point functions. The advantage of the formalism lies in the fact that, due to the self-consistent nature of the resulting equations, their solutions realize an infinite resummation of the perturbative series, leading to a more accurate description compared to ordinary perturbation theory, particularly when coupling constants are not small.

The simplest approximation of the 2PI effective action is the Hartree truncation. It leads to a momentum-independent self-energy, making the calculations particularly simple. It has been used extensively in different areas, such as chiral symmetry restoration \cite{lenaghan00,roder03} and properties of bulk viscosity \cite{dobado12}, curved spacetimes \cite{arai12}, nontopological solitons \cite{tranberg14} and superfluidity \cite{alford14,schmitt14}. The approximation represents a valuable tool if one is to look for the thermodynamic behavior of scalar theories, even though it lacks in giving information e.g., on particle lifetimes, and it also violates Goldstone's theorem. The latter can be cured by different methods, which has been also of importance and interest \cite{alford14,ivanov05,pilaftsis13}.

Renormalization of 2PI approximations has an extended body of literature. The most striking feature is the observation that the consistent cancelation of infinities cannot be achieved by equal mass and coupling counterterms \cite{hees02,borsanyi05,fejos08}. As first clarified in Ref. \cite{borsanyi05}, this property can be traced back to the fact that there are several independent representations of the propagator and the four-point function, which coincide in the full 2PI theory, but in general not in its approximations. If these quantities differ, their divergences also do; therefore, only an appropriate resummation of the perturbative series of the corresponding counterterms has to be taken into account, leading to their inequality. Furthermore, complicated group structure can also extend the number of them, which arises from various projections of the four-point function getting resummed differently; therefore, so do the projections of the counterterms themselves. We note that the splitting of the counterterms is only due to the truncation of the 2PI effective potential; given that one is able to include all diagrams, all the mass and coupling counterterms coincide. It can also be argued that the ${\cal O}(\lambda^n)$ truncation of the effective action will lead to counterterms that differ only at ${\cal O}(\lambda^{n+1})$, where $\lambda$ is the coupling constant. Without going into details, in $O(N)$-like models, the 2PI-Hartree approximation contains a single mass and three different coupling counterterms. The reader is referred to Refs. \cite{fejos08,reinosa11} for a detailed description.

Papers considering 2PI renormalization and the explicit calculation of counterterms and the effective potential itself in the broken phase of scalar theories usually assume that the condensate is homogeneous. Recently, the 2PI-Hartree approximation was used to describe a role reversal in first and second sound in a uniform superfluid \cite{alford14}, which requires the mean field to be spacetime dependent. Renormalization of this superflow-dependent condensation was also discussed, but with several ambigous points. The authors of Ref. \cite{alford14} argue that renormalizability depends on the actual renormalization conditions imposed. This peculiar statement arises from the appearance of unconventional superflow-dependent divergences found in the one-loop part of the 2PI effective potential, which seem to be able to be eliminated only when certain renormalization conditions are imposed.

In this paper, we attempt to clarify the divergence structure of the system and show that, regarding renormalizability, there is no restriction whatsoever on renormalization conditions. As it will be shown, an appropriate choice of the regularization procedure lies in the core of this statement. It will turn out that cancelation of unconventional superflow-dependent subdivergences requires the regularization to obey a certain ``phase shift symmetry'' of the quantum effective action \cite{marko14}. As a result, one needs to use a Lorentz- and translation-invariant regularization, which actually raises nontrivial questions at finite temperature. A possible resolution of these issues will also be presented.

The paper is organized as follows. In Sec. II, we introduce the model, the symmetry breaking pattern and the approximate 2PI effective potential. In Sec. III, we present the renormalization of the propagator equations and the field derivative (i.e., basically the field equation). We will put particular emphasis on differences compared to our earlier procedure described in Ref \cite{fejos08}. In Sec. IV, we show the finiteness of the effective potential explicitly and discuss and resolve the aforementioned problems of the finite-temperature calculation. Finally, in Sec. V, the reader finds some concluding remarks.

\section{Basics}

Let us consider the dynamics of a complex $\varphi$ field through the Lagrangian
\bea
\label{Eq:Lag}
{\cal L}(\varphi)=\frac12 \partial_{\mu}\varphi \partial^\mu \varphi^*-\frac{m^2}{2}\varphi \varphi^*-\frac{\lambda}{4}(\varphi \varphi^*)^2,
\eea
which displays a $U(1)$ global symmetry, with the coupling constant $\lambda>0$. We are interested in a symmetry-breaking pattern in which the condensation of the $\varphi$ field has a spacetime-dependent phase: $<\!\!\varphi\!\!>=v e^{i\psi(x)}$. In this paper, we restrict ourselves to a case in which $\partial_\mu \psi(x)=$ const., describing a uniform superfluid.  The shifted Lagrangian reads as
\begin{widetext}
\bea
\label{Eq:Lagsh}
{\cal L}(\varphi+ve^{i\psi})&=&\frac12 \partial_{\mu}\varphi \partial^\mu \varphi^*-\frac{m^2}{2}\varphi \varphi^*-\frac{\lambda}{4}(\varphi \varphi^*)^2
+\frac{v^2}{2}(\partial \psi)^2-\frac{m^2}{2}v^2-\frac{\lambda}{4}v^4
+iv(\partial_\mu \varphi^* \partial^\mu\psi e^{i\psi}-\partial_\mu \varphi \partial^\mu \psi e^{-i\psi})\nonumber\\
&-&\frac{m^2v}{2}(\varphi e^{-i\psi}+\varphi^* e^{i\psi})
-\frac{\lambda v^2}{4}(\varphi e^{-i\psi}+\varphi^* e^{i\psi})^2
-\frac{\lambda v}{2}(\varphi^*\varphi+v^2)(\varphi e^{-i\psi}+\varphi^* e^{i\psi}),
\eea
\end{widetext}
where we used the shorthand notation $(\partial \psi)^2=\partial_\mu \psi \partial^\mu \psi$. Because of the $\partial_\mu \psi$ inhomogeneity, we receive an extra term in the classical potential, coming from the kinetic term:
\bea
V[v;\psi]=\frac{m^2v^2}{2}+\frac{\lambda v^4}{4}-\frac{v^2}{2}(\partial \psi)^2.
\eea
Assuming the symmetry-breaking pattern described above, we shall build up the 2PI-Hartree effective potential of the theory and show how it is free of divergences with appropriately chosen counterterms, with particular emphasis on possible divergences caused by the appearance of the nonzero $\partial_\mu \psi$ superflow.

As mentioned in the introduction, the 2PI effective potential has two types of variables, condensates and propagators. In the usual representation, it reads as
\bea
\label{Eq:V_2PI}
V_{\textrm{2PI}}[v,{\cal G}]&=&\frac{(m^2+\delta m^2)v^2}{2}+\frac{(\lambda+\delta \lambda_4) v^4}{4}-\frac{v^2}{2}(\partial \psi)^2 \nonumber\\
&-&\frac{i}{2}\int \textrm{Tr} \hspace{0.08cm}\textrm{ln}\hspace{0.08cm}{\cal G}^{-1}-\frac{i}{2}\int \textrm{Tr} ({\cal G}_0^{-1} {\cal G}-1)+V_2,\nonumber\\
\eea 
where ${\cal G}$ and ${\cal G}_0$ are self-consistent and tree-level propagators, respectively, and $V_2$ contains all two-particle-irreducible diagrams, with vertices of the shifted Lagrangian (\ref{Eq:Lagsh}), built up by self-consistent propagators. Note that we also indicated counterterms explicitly (from now on, we shall use $m_b^2:=m^2+\delta m^2$, $\lambda_4:=\lambda+\delta \lambda_4$). The tree-level propagator around which we build up (resummed) perturbation theory corresponds to the real and imaginary parts of the {\it transformed field} $\varphi e^{i\psi}$. Its elements are
\begin{subequations}
\bea
\label{Eq:treeprop}
i{\cal G}_0^{-1}(k)_{11}&=&k^2-m_b^2+(\partial \psi)^2 - (\lambda_A+2\lambda_B) v^2, \\
i{\cal G}_0^{-1}(k)_{22}&=& k^2-m_b^2+(\partial \psi)^2 - \lambda_A v^2, \\
i{\cal G}_0^{-1}(k)_{12}&=&-2ik^\mu\partial_{\mu}\psi, \\
i{\cal G}_0^{-1}(k)_{21}&=&2ik^\mu\partial_{\mu}\psi, 
\eea
\end{subequations}
where $\lambda_A=\lambda + \delta\lambda_A$ and $\lambda_B=\lambda+\delta \lambda_B$ are {\it different} bare coupling constants corresponding to two four-index invariant tensors of the $O(2)$ group \cite{fejos08}. We remind the reader that this is due to the splitting of the $\lambda (\varphi \varphi^*)/4$ self-interaction term in (\ref{Eq:Lag}) and (\ref{Eq:Lagsh}) into two parts at bare level, if one is to calculate the tree-level propagator and/or higher loop contributions in the 2PI effective action. If we group the real and imaginary parts of $\varphi$ into a two component $\varphi_a$ vector, this splitting means
\bea
\label{Eq:split}
&&\lambda (\varphi \varphi^*)^2\equiv\nonumber\\ 
&&\equiv \frac{\lambda}{3} (\delta_{ab}\delta_{cd}+\delta_{ac}\delta_{bd}+\delta_{ad}\delta_{bc}) \varphi_a\varphi_b\varphi_c \varphi_d \nonumber \\
\longrightarrow && \quad \frac13[\lambda_A \delta_{ab}\delta_{ad}+\lambda_B (\delta_{ac}\delta_{bd}+\delta_{ad}\delta_{bc})]\varphi_a\varphi_b\varphi_c \varphi_d.
\eea
Relation (\ref{Eq:split}) basically states that different counterterms have to be associated with different invariant tensors in the interaction term, as already announced in the Introduction. Note that in the classical potential [i.e. second term on the right-hand side of (\ref{Eq:V_2PI})] no such splitting of the countercouplings is necessary; there, we used a unique $\delta \lambda_4$ counterterm.

In the Hartree approximation, $V_2$ is approximated with the double scoop diagrams,
\bea
\label{Eq:V2}
V_2&=&\frac{\lambda_A}{4} \left(\int_k \textrm{Tr}\hspace{0.08cm} {\cal G}(k)\right)^2 \nonumber\\
&+&\frac{\lambda_B}{4}\int_k \int _p \textrm{Tr} \left({\cal G}(k){\cal G}(p)+{\cal G}(k){\cal G}^T(p)\right),
\eea
where the same $\lambda_A$ and $\lambda_B$ bare couplings appeared as in the tree-level propagator. Equation (\ref{Eq:V2}) leads to a momentum-independent self-energy after differentiation with respect to ${\cal G}$, which represents the simplest approximation of the gap equations in the 2PI formalism. Note that throughout the paper the momentum integrals contain (a yet undefined) regularization, and without indicating, they are considered at some finite temperature $\tau$. 

The stationary conditions $\delta V_{\textrm{2PI}}/\delta {\cal G}=0$, $\partial V_{\textrm{2PI}}/\partial v=0$ lead to propagator and field equations. From the former, we get
\bea
\label{Eq:gapeqorig}
i{\cal G}^{-1}&=&i{\cal G}_0^{-1}\nonumber\\
&-&\lambda_A\int_k \textrm{Tr} {\cal G}(k)-\lambda_B\left(\int_k {\cal G}(k)+\int {\cal G}^T(k)\right),
\eea
while the field derivative reads as
\bea
\label{Eq:EoS}
\frac{\partial V_{\textrm{2PI}}}{\partial v}&=&v\Big(m_b^2+(\lambda +\delta\lambda_4) v^2-(\partial \psi)^2\nonumber\\
&+& (\lambda_A+2\lambda_B) \int_k {\cal G}_{11}(k) + \lambda_A \int_k {\cal G}_{22}(k)\Big).
\eea
In what follows, we shall perform renormalization on both (\ref{Eq:gapeqorig}) and (\ref{Eq:EoS}). Note that it is not necessary to require the field derivative to vanish; its expression has to be renormalizable for arbitrary values of the background field, once the solution of ${\cal G}$ is exploited. This statement does not hold for $\delta V_{\textrm{2PI}}/\delta {\cal G}$, and in (\ref{Eq:gapeqorig}), we deal with the propagator equation itself.

\section{Renormalization}

In the following, we adopt the renormalization procedure developed in Refs. \cite{fejos08,fejos09}. This is based on a scheme in which the divergence structure of a given loop integral is obtained by expanding its integrand around an auxiliary propagator $G_0(k)=i/(k^2-M_0^2)$ and identifying divergences via the zero-temperature quantities:
\begin{subequations}
\label{Eq:Td}
\bea
T_d^{(2)}&:=&\int_k^{\tau=0} G_0(k), \\
T_d^{(0)}&:=&-i\int_k^{\tau=0} G_0^2(k),
\eea
\end{subequations}
where $M_0$ plays the role of the renormalization scale. Furthermore, we also define
\begin{subequations}
\label{Eq:Td_2}
\bea
T_d^{(2),\mu\nu}&:=&-4\int_k^{\tau=0} k^\mu k^\nu G_0^2(k)\Big|_{\textrm{div}}, \\
T^{(0),\mu\nu}_{d}&:=&-4\int_k^{\tau=0} k^\mu k^\nu G_0^3(k)\Big|_{\textrm{div}}.
\eea
\end{subequations}
These integrals will appear in the divergence analysis, and they can be expressed through (\ref{Eq:Td}) (see the Appendix). Our procedure heavily relies on the fact that overall divergences cannot depend {\it explicitly} on the temperature; therefore, they can be defined through zero-temperature integrals. Note that {\it implicit} temperature-dependent subdivergences via the masses and/or the superflow might appear, and they have to be taken care of separately.

First, we discuss the renormalization of the propagator equation (\ref{Eq:gapeqorig}). Let us define the tadpole integrals as
\begin{subequations}
\bea
T(M_1;\psi):=\int_k {\cal G}_{11}(k), \\
T(M_2;\psi):=\int_k {\cal G}_{22}(k),
\eea
\end{subequations}
where ${\cal G}$ is the self-consistent propagator matrix, already introduced in the previous subsection. With the assumption of the form
\bea
\label{Eq:propass}
i{\cal G}^{-1}=
\left( \begin{array}{cc}
k^2-M_1^2+(\partial \psi)^2 & \hspace{-0.25cm} -2ik^\mu\partial_{\mu}\psi \\
2i k^{\mu}\partial_{\mu}\psi & \hspace{-0.25cm} k^2-M_2^2+(\partial \psi)^2  \\
\end{array} \right),
\eea
(\ref{Eq:gapeqorig}) leads to the following equations for the diagonal elements:
\begin{subequations}
\label{Eq:gapeq}
\bea
\label{Eq:gapeq1}
M_1^2&=&m_b^2+(\lambda_A+2\lambda_B)v^2 \nonumber\\
&+& (\lambda_A+2\lambda_B) T(M_1;\psi) + \lambda_A T(M_2;\psi), \\
\label{Eq:gapeq2}
M_2^2&=&m_b^2+\lambda_A v^2 \nonumber\\
&+& (\lambda_A+2\lambda_B) T(M_2;\psi) + \lambda_A T(M_1;\psi).
\eea
\end{subequations}
Note that, with (\ref{Eq:propass}) the off-diagonal elements of (\ref{Eq:gapeqorig}) are fulfilled automatically, since the corresponding integrands of the tadpoles are odd under the transformation $k\rightarrow -k$, and therefore their integrals give zero. 

Following the route of Ref. \cite{fejos08}, we now have to analyze the sub- and overall divergences of the tadpole integrals appearing on the right-hand sides of (\ref{Eq:gapeq}). Because of the presence of a nonzero superflow, this procedure changes compared to the analysis performed in Refs. \cite{fejos08} and \cite{alford14}. After inverting (\ref{Eq:propass}), we get
\begin{subequations}
\label{Eq:G_sol}
\bea
{\cal G}(k)_{11}&=&\frac{i}{k^2-M_1^2+(\partial \psi)^2-\frac{(2k^\mu\partial_\mu \psi)^2}{k^2-M_2^2+(\partial \psi)^2}}, \\
{\cal G}(k)_{22}&=&\frac{i}{k^2-M_2^2+(\partial \psi)^2-\frac{(2k^\mu\partial_\mu \psi)^2}{k^2-M_1^2+(\partial \psi)^2}}, \\
{\cal G}(k)_{12}&=&-2ik^\mu\partial_{\mu}\psi \cdot G(k), \\
{\cal G}(k)_{21}&=&2ik^\mu\partial_{\mu}\psi \cdot G(k). 
\eea
\end{subequations}
where $G(k)=i/[(k^2-M_1^2+(\partial \psi)^2)(k^2-M_2^2+(\partial \psi)^2)-(2k^\mu \partial_\mu \psi)^2]$. Using (\ref{Eq:G_sol}), the tadpoles read as
\begin{subequations}
\bea
\!\!\!\!\!\!\!\!\!\!\!\!T(M_1;\psi)=\int_k \frac{i}{k^2-M_1^2+(\partial \psi)^2-\frac{(2k^\mu\partial_\mu \psi)^2}{k^2-M_2^2+(\partial \psi)^2}}, \\
\!\!\!\!\!\!\!\!\!\!\!\!T(M_2;\psi)=\int_k \frac{i}{k^2-M_2^2+(\partial \psi)^2-\frac{(2k^\mu\partial_\mu \psi)^2}{k^2-M_1^2+(\partial \psi)^2}}. 
\eea
\end{subequations}
After a short calculation, for the divergent parts we get
\begin{subequations}
\label{Eq:divgap}
\bea
T(M_1;\psi)|_{\textrm{div}}&=&T_d^{(2)}+(M_1^2-\partial_\mu \psi \partial^\mu \psi-M_0^2)T_d^{(0)}\nonumber\\
&+&\partial_\mu \psi \partial_\nu \psi\cdot T^{(0),\mu\nu}_{d}, \\
T(M_2;\psi)|_{\textrm{div}}&=&T_d^{(2)}+(M_2^2-\partial_\mu \psi \partial^\mu \psi-M_0^2)T_d^{(0)}\nonumber\\
&+&\partial_\mu \psi \partial_\nu \psi\cdot T^{(0),\mu\nu}_{d}.
\eea
\end{subequations}
In the Appendix, it is shown that $T_d^{(0),\mu\nu}=g^{\mu\nu}T_d^{(0)}$, and therefore the subdivergences related to the superflow cancel. Note that this is a regularization-dependent statement. Nevertheless, as long as it obeys Lorentz invariance, the above relation remains true. We will come back to this issue later, but at this point, one concludes that there are no counterterms that need to be introduced corresponding to the superflow. 

To obtain the mass and coupling countertems, we revisit the ``one-step'' renormalization described in Ref. \cite{fejos08}, i.e., we {\it a priori} assume the existence of finite versions of (\ref{Eq:gapeq}), and insert the finite masses obtained this way to the right-hand side of the unrenormalized equations. The finite gap equations are
\begin{subequations}
\label{Eq:finM}
\bea
M_1^2=m^2&+&3\lambda v^2 \nonumber\\
&+& \lambda  T_F(M_2;\psi) + 3\lambda T_F(M_1;\psi), \\
M_2^2=m^2&+&\lambda v^2 \nonumber\\
&+&\lambda T_F(M_1;\psi) + 3\lambda T_F(M_2;\psi),
\eea
\end{subequations}
where $T_F(M_i;\psi)\equiv T(M_i;\psi)-T_{\textrm{div}}(M_i;\psi)$ [$i=1,2$]. Requiring the overall, the $v^2$, and the tadpole- (and therefore environment-) dependent subdivergences to vanish independently, one arrives at six conditions for $\delta \lambda_A$ and $\delta \lambda_B$ and two for $\delta m^2$. Only three of these relations are independent, and one recovers the results of Ref. \cite{fejos08}:
\begin{subequations}
\bea
\delta \lambda_B&=&-2\lambda T_d^{(0)}\frac{\lambda}{1+2\lambda T_d^{(0)}}, \\
\delta \lambda_A&=&-2\lambda T_d^{(0)}\frac{3\lambda+\delta\lambda_B}{1+2\lambda T_d^{(0)}}, \\
\delta m^2&=&-2(\lambda_A+\lambda_B)\left[T_d^{(2)}+(m^2-M_0^2)T_d^{(0)}\right].
\eea
\end{subequations}

Now, we turn to the field derivative $\partial V_{\textrm{2PI}}/\partial v$ (which also leads to the equation of state when one searches for its stationary point). Comparing (\ref{Eq:EoS}) with (\ref{Eq:gapeq1}), we see that we have to require $\delta \lambda_4=\delta\lambda_A+2\delta\lambda_B$ to cancel the divergences. The finite expression reads as
\bea
\frac{\partial V_{\textrm{2PI}}}{\partial v}=v\left(M_{1}^2-2\lambda v^2-(\partial \psi)^2\right).
\eea

\section{Effective potential}

The one-particle-irreducible (1PI) effective potential (up to a constant) can be obtained by substituting the solution of the propagator equations into $V_{\textrm{2PI}}$. In this section, we show that it is finite with the counterterms already determined, and all superflow-dependent divergences get eliminated, if the regularization procedure on top of Lorentz symmetry also obeys translation invariance.

Let us first start with (\ref{Eq:gapeqorig}). Multiplying both sides with ${\cal G}(p)/2$, taking the trace and integrating over $p$, we get the following useful relation (valid only for the solution of the propagator equation):
\bea
\label{Eq:G-id}
&&\frac{i}{2}\int_p \textrm{Tr}[{\cal G}_0^{-1}(p){\cal G}(p)-1]=\nonumber\\
&&\frac{\lambda_A}{2}\left[\int_k \textrm{Tr}\hspace{0.08cm}{\cal G}(k)\right]^2+\frac{\lambda_B}{2}\int_k\int_p \textrm{Tr}\left[{\cal G}(k)[{\cal G}(p)+{\cal G}^T(p)]\right].\nonumber\\
\eea
If we make use of the identity (\ref{Eq:G-id}) in $V_{\textrm{2PI}}$ [see Eq. (\ref{Eq:V_2PI})], then the simplified expression of $V_{\textrm{1PI}}$ can be obtained,
\bea
\label{Eq:1PI2PI}
V_{\textrm{1PI}}[v]&=&\frac{m_b^2}{2}v^2+\frac{\lambda_4}{4}v^4-\frac{v^2}{2}\partial_{\mu}\psi \partial^{\mu}\psi-\frac{i}{2}\int_k \textrm{Tr}\hspace{0.08cm}\textrm{ln}\hspace{0.08cm}{\cal G}^{-1}(k)\nonumber\\
&-&\frac{\lambda_B}{4}\int_k\int_p \textrm{Tr}\left[{\cal G}(k){\cal G}(p)+{\cal G}(k){\cal G}^T(p)\right]\nonumber\\
&-&\frac{\lambda_A}{4}\left(\int_k \textrm{Tr}\hspace{0.08cm} {\cal G}(k)\right)^2-N,
\eea
where $N$ is a normalization factor to be determined later, which ensures that at zero field and temperature the effective potential is zero. Note that in (\ref{Eq:1PI2PI}) the propagators should not be considered as variables but substituted solutions of (\ref{Eq:gapeqorig}). After calculating the traces, we get
\bea
\label{Eq:1pipot}
\!\!\!\!\!\! V_{\textrm{1PI}}[v]&=&\frac{m_b^2}{2}v^2+\frac{\lambda_4}{4}v^4-\frac{v^2}{2}\partial_{\mu}\psi \partial^{\mu}\psi+L(M_1,M_2;\psi)\nonumber\\
&-&\frac{\lambda^A+\lambda^B}{4}\Big(T(M_1;\psi)+T(M_2;\psi)\Big)^2\nonumber\\
&-&\frac{\lambda^B}{4}\Big(T(M_1;\psi)-T(M_2;\psi)\Big)^2-N,
\eea
where
\bea
L(M_1,M_2;\psi)&:=&-\frac{i}{2}\int_k \textrm{log}\Big[(k^2-M_1^2+\partial_{\mu}\psi \partial^{\mu}\psi)\nonumber\\
&\times&(k^2-M_2^2+\partial_{\mu}\psi \partial^{\mu}\psi)-(2k^\mu\partial_\mu \psi)^2\Big]\nonumber\\
\eea
is the remaining trace-log piece of the one-loop part [fourth term on the right-hand side of (\ref{Eq:V_2PI})].

The divergence structure of the tadpoles is already known from the previous section, and now we have to calculate $L(M_1,M_2;\psi)|_{\textrm{div}}$. The scheme we use is the same as in the previous subsection: we expand the propagators of the integrand around the auxiliary propagator $G_0(k)$ and identify the divergent terms through its zero-temperature integrals (\ref{Eq:Td}) and (\ref{Eq:Td_2}). First, we separate a quartic divergence via the term $L(M_0,M_0;\psi)$ and then identify the rest, which are all quadratic and logarithmic. We arrive at
\begin{widetext}
\bea
L(M_1,M_2;\psi)|_{\textrm{div}}&=&L^{\tau=0}(M_0,M_0;\psi)
+\left(M_1^2+M_2^2-2M_0^2-2(\partial \psi)^2\right)\frac{T_d^{(2)}}{2}\nonumber\\
&+&\left[\left(M_1^2-M_0^2-(\partial \psi)^2\right)^2+\left(M_2^2-M_0^2-(\partial \psi)^2\right)^2\right]\frac{T_d^{(0)}}{4}
+(\partial\psi)^2 T_d^{(2)}-(\partial\psi)^4 \frac{T_d^{(0)}}{2}\nonumber\\
&+&\partial_\mu\psi\partial_\nu\psi\left[\frac{T_d^{(2),\mu\nu}}{2}+\left(M_1^2+M_2^2-2M_0^2-2(\partial \psi)^2\right)\frac{T_d^{(0),\mu\nu}}{2}\right] 
-\partial_\mu\psi\partial_\nu\psi \left[\frac{T_d^{(2),\mu\nu}}{2}-(\partial \psi)^2T_d^{(0),\mu\nu}\right].\nonumber\\
\eea
\end{widetext}
With the use of the expressions of divergent quantities $T_d^{(0),\mu\nu}$ and $T_d^{(2),\mu\nu}$, which are given in the Appendix, we realize that all $\psi$ dependence cancels, except the first term on the right-hand side.
\bea
\label{Eq:tracelogdiv}
L(M_1,M_2;\psi)|_{\textrm{div}}&=&L^{\tau=0}(M_0,M_0;\psi)\nonumber\\
&+&(M_1^2+M_2^2-2M_0^2)\frac{T_d^{(2)}}{2}\nonumber\\
&+&\Big((M_1^2-M_0^2)^2+(M_2^2-M_0^2)^2\Big)\frac{T_d^{(0)}}{4}.\nonumber\\
\eea

The normalization factor $N$ in (\ref{Eq:1pipot}) is determined by the condition that at zero temperature, $V^{\tau=0}_{1PI}(v=0)=0$. Let us denote the solution of the gap equations (\ref{Eq:gapeq}) by $M^2$ at zero field and temperature (the two equations coincide in this case),
\bea
\label{Eq:gapeqsymm}
M^2=m_b^2 + 2(\lambda_A+\lambda_B) T^{\tau=0}(M;\psi).
\eea
where
\bea
T^{\tau=0}(M;\psi):&=&i\int_k^{\tau=0} \Bigg(k^2-M^2+\partial_{\mu}\psi \partial^{\mu}\psi\nonumber\\
&-&\frac{(2k_\mu \partial^\mu \psi)^2}{k^2-M^2+\partial_{\mu}\psi \partial^{\mu}\psi}\Bigg)^{-1}.
\eea
The normalization factor is then
\bea
\label{Eq:N}
N=L^{\tau=0}(M,M;\psi)-(\lambda^A+\lambda^B)T^{\tau=0}(M;\psi).
\eea
The 1PI effective potential (at finite temperature in general) is therefore
\bea
\label{Eq:1piunren}
V_{\textrm{1PI}}[v]&=&\frac{m_b^2}{2}v^2+\frac{\lambda_4}{4}v^4-\frac{v^2}{2}\partial_{\mu}\psi \partial^{\mu}\psi\nonumber\\
&+&L(M_1,M_2;\psi)-L^{\tau=0}(M,M;\psi)\nonumber\\
&-&\frac{\lambda^A+\lambda^B}{4}\Big(T(M_1;\psi)+T(M_2;\psi)\Big)^2\nonumber\\
&-&\frac{\lambda^B}{4}\Big(T(M_1;\psi)-T(M_2;\psi)\Big)^2\nonumber\\
&+&(\lambda^A+\lambda^B)\left(T^{\tau=0}(M;\psi)\right)^2.
\eea
We have seen in the previous subsection that, if the regularization obeys Lorentz invariance, the tadpoles have no superflow-dependent overall divergence, but one still might be worried about the same type of divergences in $L(M_1,M_2;\psi)$ [see the first term on the right-hand side of (\ref{Eq:tracelogdiv})] and therefore also about the applied subtractions of $N$, which should be environment independent. The term in question can be also written in the form of
\bea
\label{Eq:LM0}
L^{\tau=0}(M_0,M_0;\psi)&=&-\frac{i}{2}\int_k^{\tau=0} \log\Big[\left((k-\partial \psi)^2-M_0^2\right)\nonumber\\
&\times&\left((k+\partial \psi)^2-M_0^2\right)\Big],
\eea
where, if the regularization does not break translation invariance, we can shift the integration momenta separately to get
\bea
L^{\tau=0}(M_0,M_0;\psi)&=&-i\int_k^{\tau=0} \log (k^2-M_0^2)\nonumber\\
&\equiv& L^{\tau=0}(M_0,M_0;0),\nonumber\\
\eea
which is $\psi$ independent. The same argument leads to relations $L^{\tau=0}(M,M;\psi)=L^{\tau=0}(M,M;0)$ and $T^{\tau=0}(M;\psi)=T^{\tau=0}(M;0)$, and therefore $N$ is also $\psi$ independent. (Note that, for example, any type of cutoff regularization explicitly breaks translation invariance, and in this case, depending on the validity of Lorentz invariance the tadpoles might not, but (\ref{Eq:LM0}) does contain a $\psi$ dependent overall divergence.) The symmetry behind this $\psi$ independence is the invariance of the zero temperature 1PI effective action (based on formal considerations) \cite{marko14},
\bea
\Gamma_{1PI}^{\tau=0}[\varphi e^{-i\alpha x};\partial_\mu\psi]=\Gamma_{1PI}^{\tau=0}[\varphi;\partial_\mu \psi-\alpha_\mu],
\eea
which shows that at zero field expectation value $\psi$ is only a spurious field having no physical relevance. Nevertheless, if one chooses a regularization that breaks this invariance explicitly, then $\psi$-dependent divergences can and will be generated.

We still have to check the cancelation of environment-dependent subdivergences in (\ref{Eq:1piunren}), which appear via the masses $M_1$ and $M_2$.
The easiest way to show that (\ref{Eq:1piunren}) is finite is to follow the route of Ref. \cite{marko13}. One exploits the unrenormalized equations (\ref{Eq:gapeq}) and (\ref{Eq:gapeqsymm}), expresses the tadpoles, and then substitutes them into (\ref{Eq:1piunren}). After a short calculation, one arrives at the finite expression
\bea
\label{Eq:1piren}
V_{1PI}[v]&=&M_{1}^2\frac{v^2}{2}-\lambda \frac{v^4}{4}-\frac{v^2}{2}\partial_\mu \psi \partial^\mu \psi\nonumber\\
&+&L_F(M_{1},M_{2};\psi)-L_F^{\tau=0}(M,M;\psi)\nonumber\\
&+&\Big(4m^2(M_{1}^2+M_{2}^2-2M^2)-(M_{1}^2-M_{2}^2)^2\nonumber\\
&-&2(M_{1}^4+M_{2}^4-2M^4)\Big)/32\lambda, 
\eea
where 
\bea
L_F(M_{1},M_{2};\psi)=L(M_{1},M_{2};\psi)-L(M_{1},M_{2};\psi)|_{\textrm{div}},\nonumber\\
\eea
and correspondingly
\bea
L^{\tau=0}_F(M,M;\psi)=L^{\tau=0}(M,M;\psi)-L(M,M;\psi)|_{\textrm{div}}.\nonumber\\
\eea

Equation (\ref{Eq:1piren}) shows the explicit finiteness of the effective potential and that it is properly normalized. Note that, depending on actual model parameters (and possibly on the superflow itself), it might not be possible to access $v=0$ at zero temperature (due to the disappearance of the solution of the propagator and field equations). In this case, one has to choose another subtraction point for defining the normalization factor $N$, e.g., the minimum of the effective potential.

Finally, let us discuss an ambiguous point of the procedure described above, appearing at finite temperature. We saw that at any temperature $\tau$ counterterms (defined at zero temperature) render all sub- and overall divergences finite, but we have not yet addressed the question of how the demands of regularization (i.e. Lorentz and translation invariance) and a finite-temperature calculation can be accommodated. This issue is nontrivial due to the following. 

When calculating the effective potential, one has to perform (e.g. in imaginary time formalism) Matsubara sums in the trace-log term $L(M_1,M_2;\psi)$ and also in the tadpoles $T(M_1;\psi)$, $T(M_2;\psi)$. These summations can be done analytically, leading each term to a three-dimensional momentum integral. But after this step, Lorentz invariance is immediately broken, and if one cuts the momentum integral with a UV cutoff, translational invariance will also be lost. As discussed in the previous subsections, this leads to superflow-dependent divergences in both the one-loop and tadpole integrals. One has two choices at this point: {\bf 1}) keep track of these divergences and subtract them by hand, since these are only related to a ``bad'' choice of regularization, or {\bf 2}) choose instead a Lorentz- and translation-invariant regularization even at finite temperature. Let us follow the second choice.

Even though, through the implicit temperature dependence of the masses, one cannot define finite-temperature and vacuum parts of the diagrams properly (since $\tau$ will remain implicitly in the latter one), it is always possible to separate the explicit temperature dependence from the implicit one. The importance of this lies in the fact that only the former, ``vacuum'' parts contain overall divergences, and therefore only these need to be regularized. In other words, only in these terms do we need to apply a Lorentz- and translation-invariant regularization. For example, after performing the Matsubara sum, $L(M_{1},M_{2};\psi)$ reads as
\bea
\label{Eq:ring}
L(M_{1},M_{2};\psi)=\sum_{i=1,2} \int &&\!\!\!\!\!\!\frac{d^3k}{(2\pi)^3}\Big(\omega_i({\bf k})/2 \nonumber\\
&+&\tau\textrm{ln}(1-e^{\omega_i({\bf k})/\tau})\Big),
\eea
where $\omega_i({\bf k})$ [$i=1,2$] is the energy of an eigenmode (determined by the zeros of the propagator determinant). In (\ref{Eq:ring}), only zero-point fluctuations (first term in the bracket) diverge, but as mentioned already, in its current form, it is not suitable for avoiding the appearance of superflow-dependent divergences. The way out is to rewrite {\it only} the zero-point fluctuations into their $\tau=0$ original form (i.e., before performing the Matsubara sum) or to actually {\it define} the finite temperature $L(M_{1},M_{2};\psi)$ as
\bea
\label{Eq:ring2}
L(M_{1},M_{2};\psi)&=&\int_k^{\tau=0} \textrm{log}\Big((k^2-M_{1}^2+\partial_\mu \psi \partial^\mu \psi)\nonumber\\
&\times&(k^2-M_{2}^2+\partial_\mu \psi \partial^\mu \psi)-(2k^\mu \partial_\mu\psi)^2\Big)\nonumber\\
&+&\sum_{i=1,2}\int \frac{d^3k}{(2\pi)^3}\tau\textrm{ln}(1-e^{\omega_i({\bf k})/\tau}),
\eea
instead of (\ref{Eq:ring}). In the first term of the right-hand side of (\ref{Eq:ring2}) now we can apply an appropriate regularization, while the second term is completely finite. We therefore solved the problem: we obtained a form in which the divergent integral can be Lorentz and translation invariant, and at the same time, it also describes the finite-temperature behavior. 

The same kind of procedure has to be applied also to every tadpole integral: after separating the vacuum from the explicit temperature-dependent part, one rewrites the former as a Lorentz-invariant integral, and defines its divergence using a Lorentz- (or Euclidean after Wick rotation) and translation-invariant regularization, which leads eventually to the disappearance of all superflow-dependent divergences. Nevertheless, as already mentioned, if one is to use a regularization breaking the previous properties, then new, superflow-dependent counterterms have to be added to the Lagrangian. An analysis of this type is beyond the scope of the paper, but since the procedure described here works without any restrictions (even at finite temperature), we do not feel the necessity of such an approach.

\section{Conclusions}

In this paper, we investigated whether the 2PI-Hartree approximation is renormalizable in the broken phase with a nonzero superflow in a $U(1)$ symmetric scalar theory. Somewhat contrary to the findings of Ref. \cite{alford14}, we argued that with the counterterms already determined in Ref. \cite{fejos08} there is no ambiguity of the effective potential; it is finite and well defined at all renormalization scales ($M_0$). We have found two main differences compared to the analysis of Ref. \cite{fejos08}: {\bf 1}) in the effective potential, one-loop and tadpole integrals might contain divergences related to the superflow, but if {\bf 2}) a Lorentz- and translation-invariant (but otherwise completely arbitrary) regularization is used, these do not appear at all. Concerning the finite-temperature treatment, we proposed to separate the loop integrals as sums of the explicit and implicit temperature-dependent parts and rewrite (or actually define) the former one using a Lorentz- and translation-invariant regularization, in order to avoid the appearance of environment-dependent divergences.

The ambiguous findings of Ref. \cite{alford14} are due to the incompleteness of the divergence analysis of the 2PI effective potential. On the one hand, the authors miss that the double scoop diagrams might lead to superflow-dependent divergences, if the regularization breaks Lorentz invariance, and on the other hand, they skip the analysis of the sensitivity of the divergence structure of the one-loop part with respect to the regularization used. Since they ultimately neglect all the vacuum parts, it would be interesting to see how and in what regime these terms were of importance from the view of the solution of the coupled propagator and field equations. The renormalization method and our explicitly finite representation of the effective potential given here would allow one to perform such an investigation in a straightforward way.

\section*{Acknowledgements}
The author thanks Urko Reinosa for drawing attention on the symmetry property of the effective action and also for useful comments concerning the manuscript. The careful reading of the manuscript by Zsolt Sz\'ep is also greatly acknowledged, together with discussions with Gergely Mark\'o. This work was supported by the Foreign Postdoctoral Research program of RIKEN.

\makeatletter
\@addtoreset{equation}{section}
\makeatother 
\renewcommand{\theequation}{A\arabic{equation}} 
\appendix
\section{Divergent integrals}

In the Appendix we calculate the divergent quantities of (\ref{Eq:Td_2}). Assuming that the regularization does not break Lorentz symmetry, both $T_d^{(0),\mu\nu}$ and $T_d^{(2),\mu\nu}$ have to be proportional to $g^{\mu\nu}$, since this is the only two-index tensor that is Lorentz invariant. For $T_d^{(0),\mu\nu}$, we have
\bea
T_d^{(0),\mu\nu}=g^{\mu\nu}i\int_k^{\tau=0} \frac{k^2}{(k^2-M_0^2)^3}\bigg|_{\textrm{div}}.
\eea
Adding and subtracting $M_0^2$ in the numerator, we immediately see that
\bea
T_d^{(0),\mu\nu}=g^{\mu\nu} T_d^{(0)}.
\eea
The other integral is
\bea
T_d^{(2),\mu\nu}=\int_k^{\tau=0} \frac{4k^\mu k^\nu}{(k^2-M_0^2)^2}\bigg|_{\textrm{div}}.
\eea
Similarly to $T_d^{(0),\mu\nu}$, we exploit Lorentz symmetry and write
\bea
T_d^{(2),\mu\nu}=\int_k^{\tau=0} \frac{g^{\mu\nu}k^2}{(k^2-M_0^2)^2}\bigg|_{\textrm{div}},
\eea
which is
\bea
T_d^{(2),\mu\nu}=g^{\mu\nu}(T_d^{(2)}+M_0^2T_d^{(0)}).
\eea

\end{document}